# NavMarkAR: A Landmark-based Augmented Reality (AR) Wayfinding System for Enhancing Spatial Learning of Older Adults

ZHIWEN QIU*, Cornell University, USA

MOJTABA ASHOUR, Cornell University, USA

XIAOHE ZHOU, Cornell University, USA

SALEH KALANTARI, Cornell University, USA

Wayfinding in complex indoor environments can be challenging for older adults due to natural declines in navigational and spatial-cognition abilities. To assist in improving older adults' spatial navigation skills, we developed NavMarkAR, an augmented reality (AR) navigation system using smart-glasses to provide landmark-based navigational guidance. This project fills an important gap in design research and practice, as there have been few prior efforts to evaluate the cognitive implications of AR-aided navigational systems. We conducted an initial usability test with 6 participants, followed by design refinement of the prototype and then a more in-depth study with 32 participants, who were asked to use the NavMarkAR system while completing navigation tasks in a university building. We found that participants using NavMarkAR completed the wayfinding tasks faster and developed more accurate cognitive maps. In future research we will study the long-term cognitive skill maintenance effects of using such navigational aids.

## 1 INTRODUCTION

In the modern world, many individuals spend up to 95% of their time indoors [12]. We often face challenges when navigating such interior environments, particularly within more complex structures such as hospitals, transit hubs, and educational facilities [13]. Thus, effective navigation within these settings becomes a crucial aspect of facilitating daily activities. These challenges can be particularly pronounced for older adults, since our spatial navigation abilities tend to deteriorate as we grow older [2, 65]. Researchers have robustly demonstrated that older adults get lost more often than younger people when attempting to navigate complex facilities [43]. Such navigational difficulties can hinder older adults' autonomy, mobility, social participation, and safety [2], and significantly affect their perceived quality of life [31].

Navigation involves multiple cognitive, sensory, and motor functions [56]. It requires processing various spatial cues, engaging executive control functions, and constructing mental representations of the environment that encompass aspects such as distances, directions, spatial layouts, and the visual characteristics of a space [19, 41]. These processes are believed to rely on two primary spatial frames of reference: egocentric and allocentric. In an egocentric frame of reference, locations are encoded relative to the navigator, whereas in an allocentric frame of reference, they are independent of the navigator's location [61]. Navigation that draws from an allocentric reference frame activates greater brain activity in areas responsible for spatial orientation, and is associated with converting egocentric perspective into a map-like understanding [42]. Allocentric strategies thus enhance the ability to navigate unexplored areas and maintain spatial memory more effectively when compared to egocentric methods [6].

There is a growing body of research indicating that cognitive declines associated with aging tend to reduce the effectiveness of allocentric navigational strategies, with less impact on egocentric strategies [17]. However, it is widely recognized that successful navigation often requires the integration of both frameworks [54]. Therefore, navigational aids for older adults can benefit from focusing on improving allocentric processing and its integration with the

Authors' Addresses: Zhiwen Qiu, Cornell University, Ithaca, NY, USA, zq76@cornell.edu; Mojtaba Ashour, Cornell University, Ithaca, NY, USA; Xiaohe Zhou, Cornell University, Ithaca, NY, USA; Saleh Kalantari, Cornell University, Ithaca, NY, USA.

egocentric reference. Some evidence has shown that the choice of primary strategy (egocentric vs. allocentric) depends on the availability of environmental cues [36], such as geometric cues about the features of a space (e.g., position of a building's walls) and landmark cues that can serve as stable and prominent points of reference [14]. The selection of these reference features appears to be highly individualistic; for example, one person may consider the wall color of a space as a landmark and point of reference, while another person may ignore the color and focus on the function of that space (e.g., a seating area, or the space where everyone gathers for Friday events). For this reason, an effective navigation tool should provide a wide range of associative information about landmarks to broaden the choices for points of reference and provide the user with multiple association possibilities. The ultimate goal of this strategy is to prompt the formation of better allocentric cognitive maps, particularly in less-familiar environments.

Most existing navigation systems (e.g., Google Maps) create direction-giving scenarios in which users may feel little need for spatial learning. In other words, users are passively supplied with directional information without having to interact with their environment in any meaningful way [40]. This passive navigation approach can potentially exacerbate the decline of spatial skills and discourage the development of cognitive maps [40]. An augmented reality (AR) approach to navigational guidance—in which information is layered over the visible environment through the use of real-time video or transparent displays such as smart glasses—can improve upon this paradigm by encouraging users to engage more directly with the features of the surrounding built environment. AR superimposes the relevant navigational information onto the real-world view rather than replacing the existing view or distracting the navigator from their surroundings [3]. Researchers have found that AR approaches result in less mental workload compared to presenting navigational information on a non-transparent display [62]; such decreased workload may in turn aid in the effective development of cognitive maps [34].

In the current project, our goal was to investigate the cognitive implications of AR navigation systems for older adults, with the understanding that such technologies may assist in the formation of cognitive maps and the maintenance of spatial skills. The following two research questions were formulated to guide the study:

**RQ1.** Will a landmark-based AR navigational support system enhance the development of cognitive maps while maintaining wayfinding performance in an older adult population?

**RQ2.** What are the preferences of older adults regarding the multimodal interactions and user experience in landmark-based AR navigational support systems?

To answer these questions, we developed a novel AR landmark-based navigational assistance prototype, which is called "NavMarkAR," with the overt goal of enhancing spatial learning abilities and cognitive map development among older adults. We then conducted two user studies: the first examined the usability of NavMarkAR, while the second evaluated the efficacy of the navigation system during wayfinding. This research makes several notable contributions to the field. First, we designed and implemented a novel indoor navigation support tool specifically targeted toward older adults, focusing on human–environment interactions. Unlike other navigation tools that rely primarily on egocentric approaches, NavMarkAR allows older adults to effectively acquire and strengthen allocentric spatial knowledge by broadening the information associated with real-world landmarks and prompting engagement with the environment [51]. Second, the empirical research provides a robust evaluation of the system's effectiveness as well as important feedback from a user experience perspective with the elderly population. These findings lead to important design implications for future AR navigational interfaces. Conducting user studies afforded an opportunity for older adults to provide their input to enhance the various features of NavMarkAR while also validating the efficacy of the system for wayfinding performance and cognitive map development.

## 2 RELATED WORK

### 2.1 Existing Approaches to Navigational Support

People may receive navigational information in numerous forms. They may be given verbal directions to a specific destination, they may look at a map, they may explore the actual physical environment, or they may use turn-by-turn (TBT) navigational applications such as Google Maps. In the realm of technological systems for navigational support, TBT applications are the most common approach. Such systems guide the user to the destination through a continuous sequence of visual or spoken turning instructions (e.g., "turn left at the next intersection"), often detailing street names and distances to the next turn [39]. While TBT navigation is efficient in directing users to their destinations, empirical evidence suggests that reliance on these methods can adversely affect spatial cognition. Ruginski and colleagues [30], for example, found that long-term GPS use was correlated with a notable decrease in the ability to conduct spatial transformations and to learn the spatial features of new environments. Similarly, Hejtmánek and colleagues [45] reported that a reliance on TBT instructions in a virtual environment diminished cognitive map formation. Other research has shown that individuals who make extensive use of TBT navigational support systems tend to remember fewer landmarks [44], and have greater difficulty reaching destinations when the assistance is not available, compared to peers who seldom use TBT tools [71]. These negative effects on spatial memory may be attributable at least in part to the need for navigators to shift perspectives and divide their attention between the navigational interfaces and their surroundings [1, 9], which discourages active environmental engagement and information-gathering [5, 72].

Recent research has highlighted the need for alternative approaches to navigational aids that will better support cognitive map development, and has shown that such approaches are feasible. Chrastil and Warren [21] demonstrated that allowing active decision-making during navigation improved participants' acquisition of topological graph knowledge and self-orientation. Huang and colleagues [25] introduced a "Potential Route Area" approach that offers users the freedom to choose their own navigational paths within an indicated geographical space, and found that such guidance was associated with better spatial knowledge acquisition compared to a TBT system. Lu and colleagues [33] likewise found that navigational aids prompting active environmental engagement significantly enhanced landmark and survey knowledge. Clemenson and colleagues [24] proposed an auditory beacon system that conveys information about distance and direction, and found that it improved spatial orientation among newcomers to an area when compared to a TBT system.

Some researchers have noted that built indoor environments tend to lack the types of prominent and information-rich landmarks that we have evolved to rely on during outdoors navigation [10]. By drawing attention to certain features of the indoor environment and overlaying them with meaning, landmark-based navigational aids can in theory transform ordinary indoor layouts into more comprehensible and memorable spaces [3, 9, 22].

### 2.2 AR Navigational Support System and Cognitive Map Development

AR navigational support systems use an overlay approach to present information about the environment to users through handheld devices or transparent head-mounted displays (HMDs). These AR systems rely on three core technological components. First, localization techniques, which identify the precise real-world position and sometimes the viewpoint of the user, are crucial for aligning the virtual information with the real environment. The Global Positioning System (GPS) is primarily used for this purpose outdoors, while indoor navigation employs diverse methods including Simultaneous Localization and Mapping (SLAM), QR code-based markers, Inertia Measurement Units (IMUs), Bluetooth beacons, or computer-vision technologies [5, 37]. Some AR applications have used spatial mapping to allow more robust tracking of user positions; this involves producing a detailed 3D representation of the environment within the technological system [48]. The second technological aspect of AR navigational support is path-generation, which is used to determine an optimal route from the user's current location to their intended destination. Algorithms such as Dijkstra [18] or A* [23] are used for this purpose. Finally, the AR system needs to be able to present visual or auditory navigational aids overlaid onto the real environment, either through the use of a transparent display screen (such as an HMD) or by adding these aids to a camera feed of the environment (e.g., on a smart phone). The navigational aids may include some combination of maps, action-signs, turning directions, and/or points-of-interest (POIs), resulting in an effective user-interface [34, 70, 78, 79].

As noted in Section 2.1, an important consideration in the development of AR navigational aids is their potential to enhance or maintain users' cognitive mapping abilities, in comparison to displays that require more divided attention between an informational text and the real-world surroundings. In theory, the transparent overlay of information onto the real-world environment could encourage more active cognitive engagement with that environment. However, there has not been much robust research conducted in this area, and the prior studies that do exist regarding AR navigational aids and cognitive map development have produced inconclusive and conflicting results. Stefanucci and colleagues [30] examined overlaid navigational beacons and street-name indicators in a virtual urban environment presented through a head-mounted display, and found positive results, with the AR information correlating to improved spatial knowledge acquisition. Liu and colleagues [10] also found positive results for spatial knowledge acquisition when using a smart-glasses display to add entirely new virtual landmarks superimposed onto a real-world indoor environment. A study on indoor environments by Zhang and colleagues [34] had mixed outcomes, with AR maps yielding improved spatial learning outcomes, but simpler AR guideposts failing to show any significant learning impacts compared to participants who navigated without aids. A few studies have suggested that AR navigational support can have negative effects on spatial learning; for example, Rehman and Cao [74] observed that although AR navigational support using Google Glasses reduced cognitive workload during outdoor wayfinding, participants who used the AR exhibited more route-retention errors than those who used paper maps. Dong and colleagues [78] found that participants using handheld AR devices paid less visual attention to their environments compared to those using 2D digital maps, leading to greater difficulties for the AR group in forming a clear memory of the route. Thus, overall, a clear consensus has not yet emerged in the research community regarding what effects (if any) AR navigational tools might have on spatial learning and cognitive map formation.

### 2.3 AR Memory Aids for Older Adults and People with Mild Cognitive Impairments

AR has previously been used to develop memory aid systems with the goal of enhancing independence and rehabilitating memory issues in older adults and cognitively impaired populations. While these AR applications are not all specifically intended to help with wayfinding success or in the formation of cognitive maps, many of them address spatial memory and are thus highly relevant to the current study. Wallace and Morris [73] proposed a memory assistance app that allowed brain-injury patients to voice-record information and then later receive their own informational reminders via Bluetooth beacons when a recorded contact was nearby. Their evaluation suggested that this tool was effective for aiding patients to recall contact names and information shared with them. Hervás and colleagues [59] designed a similar system for cognitively impaired individuals, which provided point-of-interest information with landmarks and allowed close relatives to add location-specific notes. Participants reported significantly higher satisfaction towards the usability of the system compared to a commercial GPS navigator.

There is a strong scientific basis for the general use of landmark information, or "loci method", which pairs graphical information with specific physical locations for memory enhancement among individuals with cognitive impairment [80]. Iggena and colleagues [15] evaluated the relevance of visual, vestibular, and proprioceptive input on spatial memory in participants with hippocampal dysfunction, and found that multimodal cues were most effective, potentially due to the brain's ability to compensate for navigational deficits by activating extrahippocampal brain areas. Similarly, Cogné and colleagues [50] concluded that directional arrows and salient landmarks were the most helpful visual cues for spatial memory formation among individuals with MCI. These findings suggest that in individuals whose core spatial navigation abilities have been diminished, associative approaches that activate other types of memory (for example, "the shop is located to the right of my nephew's house") may be highly effective.

For older adult wayfinders, simple and streamlined navigational guidance is often the most useful. Montuwy and colleagues [4] observed that older adults had difficulty maintaining spatial awareness when they were receiving navigational instructions via smartglasses, compared to those from bone conduction headphones and haptic wearable devices. Tang and Zhou [43] found that older adult pedestrians preferred directional cues without a virtual navigator and overview map, potentially due to high cognitive demands from complex navigational aids. Similarly, Firouzian and colleagues [7] evaluated an AR-eyeglass assistive device that used simple turn indicator lights for older adults with MCI, concluding that participants were satisfied with the simplicity of guidance for enhancing environmental engagement. Kim and Dey [67] and Schall and colleagues [47] created similar AR navigational displays for vehicle windshields to assist older adult drivers, again finding that a minimalist approach to the AR visual cues was most effective for awareness of surroundings to recall context-sensitive information without compromising performance of driving tasks.

## 3 SYSTEM DESIGN

### 3.1 Prototype Development

We developed the AR navigation assistance system using HoloLens 2 smartglasses [28] and the Unity3D graphics engine version 2020.3.48f1 [75]. The NavMarkAR user interface was implemented with the support of the Microsoft Mixed Reality Toolkit (MRTK) version 2.7.3 [53]. Designing the system required three steps, as shown in Figure 1. First, we utilized Scene Understanding SDK [66] for spatial mapping, generating a structured representation of the real-world environment with essential features such as walls, floors, and ceilings [38]. These 3D models were essential for accurate positioning of the overlaid allocentric navigational components. The Scene Understanding SDK is able to detect elements of the surroundings and correlate them with the virtual representation, facilitating seamless interactions between the virtual and real worlds and ensuring that virtual content is positioned in a realistic and context-aware manner. This includes effective occlusion dynamics, so that virtual information is not presented when the relevant landmark is blocked from the users' view by a wall or other object.

Second, we used the spatial data and the built-in HoloLens sensors to achieve precise real-time localization of user positions. This ensured that the user remains consistently and accurately situated within the virtual representation, allowing for effective tracking and visualization of their trajectory. Finally, we incorporated a user experience package from MRTK to design the wayfinding guidance. With a focus on allocentric knowledge acquisition for older adults, we prioritized the landmark-based navigation paradigm over more commonly used turn-by-turn instructions or directional indicators. In addition to creating a more intuitive and engaged navigation experience, this approach has fewer computational demands and thus increases system performance and fluidity.

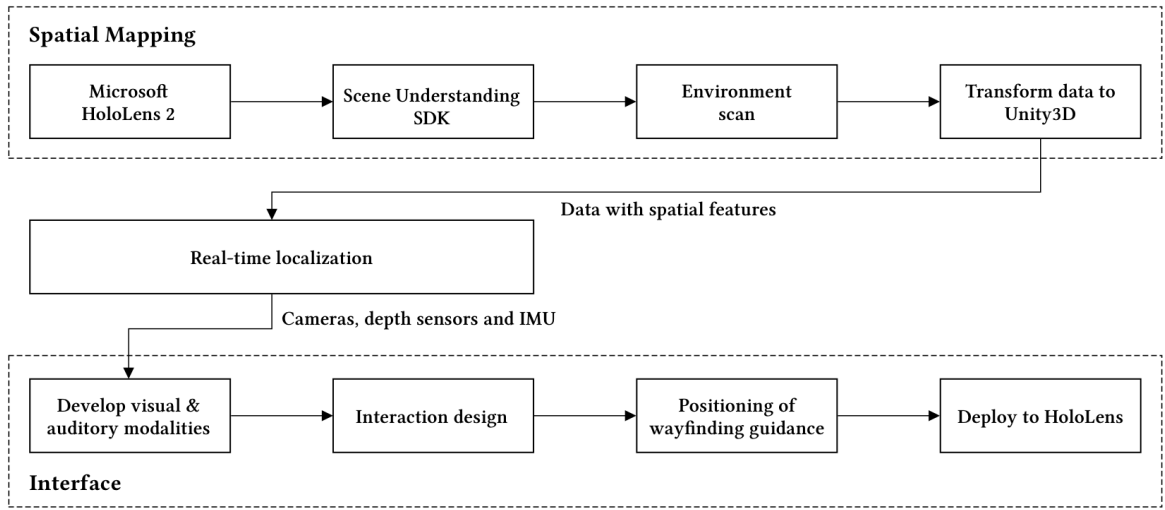

Figure 1: The system design pipeline for NavMarkAR. We first employed HoloLens with Scene Understanding SDK to scan the environment, creating a 3D representation of the space. We then used the spatial positioning features of HoloLens to pinpoint users' real-time positions. Finally, interfaces with multimodal interaction were developed to provide the wayfinding guidance.

### 3.2 Design of Wayfinding Guidance

One of the goals of NavMarkAR was to provide users with multiple types of information to draw from, but without becoming overwhelming, and while prioritizing landmark recognition. Instead of superimposing extensive elements of landmarks and directions onto the physical environment, we sought to help users connect with landmarks based on their personal interpretations for aiding in spatial knowledge acquisition. We used three types of information in the system: landmark highlights and descriptions, a 3D interactive map, and directional guidance. The system was designed to allow users to fluidly determine when these different types of guidance would appear. The wayfinding guidance was provided in both visual and auditory formats.

*3.2.1 Interface.*

To encourage user engagement with the system while preventing cognitive overload, we incorporated the pressable buttons module from MRTK [53]. This module presents virtual buttons in the AR field of view, which users can activate with their fingers (the physical/virtual interaction of these button presses is tracked via the HoloLens sensors). This feature empowered users to access navigational aids as needed, allowing them to manage and adapt the guidance to their preferences and memory retention abilities. The buttons in our system allowed users to activate the "Landmark" information and/or the "Directions" information whenever they felt necessary. When a button was pressed, both a visual and auditory announcement were triggered to affirm the interaction. To prevent unintentional activations, the system was designed to display virtual guidance only when a touchpoint was fully pressed and released. Landmark and directional information was presented in floating AR windows, which users could easily reposition by "grabbing" and "dragging" the title bars. This allowed them to customize the interface to suit their preferences and to maximize visibility. For the 3D map component, we used standard hand-tracking techniques [53]. Users could bring up the map by lifting their left hand with the palm open and facing away from their body, and then navigate through the map using hand motions. Any of these windows could be removed from the visual area by pressing a "close" button, or in the case of the map, by the user simply lowering their hand out of the visual area (Figure 2).

*3.2.2 Landmark Information.*

Inspired by the "loci" method, our navigational support system was based on selecting memorable landmarks in the environment and overlaying them with meaning to enhance participants' recall of familiar places along a route [80]. Salient landmarks, characterized by high visibility (e.g., shape, color), structural prominence (e.g., edges, boundaries), and cognitive engagement potential (e.g., personal interests, notable histories), are effective in aiding wayfinding and spatial memory [11, 16, 63]. Placing landmarks on-route or at decision points are also beneficial in route retention while navigating in indoor environments. Therefore, landmarks were selected by the researchers on the basis of their being popular locations in the building, highly visible, and in close proximity to route decision-points in our system. We also introduced a blue polyhedron icon in the AR display to signify these landmarks. The distinctive polyhedron acted as a sign to alert users that a landmark was in sight, and served to draw attention to the touchpoint where users could activate the landmark's information display. This information was presented via multisensory inputs that combined textual history and stories with pictures to increase saliency and memorability [15, 59]. Users could also activate auditory descriptions to further enrich the experience. During our empirical study in which users were asked to complete specific navigation tasks, we also used this system to signal end-points, including an auditory confirmation that announced, "Congratulations, you have arrived at the destination [3]!" The overall goal of this landmark system was to prompt users to engage more fully with the environment and learn about its notable features, thereby aiding in the development of their cognitive maps in an interesting and informative fashion.

*3.2.3 Directional Guidance.*

Directional cues were strategically programmed for key intersections. When users engaged with the interactive directional touchpoints, they were provided with a window that indicated both the direct route toward their destination and other nearby landmark points that they could reach by following alternative routes. By displaying multiple potential paths, each with relevant landmarks indicated, we sought to draw users' attention to the surrounding environment and mitigate any tendencies to favor step-by-step directions over environmental engagement. The presentation of multiple possible routes in this fashion has been previously shown to encourage spatial knowledge acquisition [25]. Each potential route was displayed with directional arrows, pictures of a nearby landmark along that route, and the landmark's name. Landmark-based auditory guidance was also included when a route was selected, such as: "Turn left when you reach the commons area."

*3.2.4 3D Interactive Map.*

The interactive map allowed users to view the entire building floorplan, with location indicators for the user's current position and for nearby landmarks [34, 54]. The purpose of including this map was to further encourage allocentric perspectives and cognitive map formation. The position of users was updated in real-time on the map as they moved through the building, providing a consistent and reliable navigation experience. However, we did not want users to depend excessively on this map for reaching their destinations, so we did not allow them to leave the map permanently visible on the AR screen. Instead, users had the choice to hold their left hands extended in front of their body to bring up the map and to keep it visible. When users lowered their hand the map disappeared, directing attention back to the surrounding environment and its features.

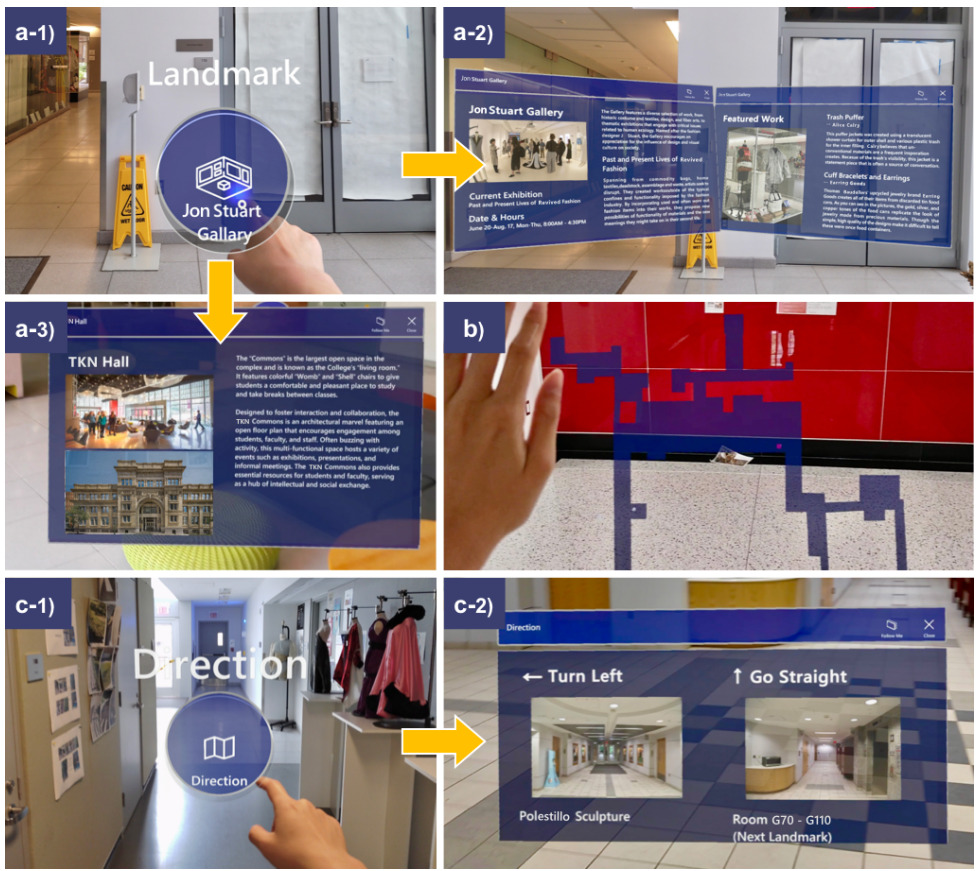

Figure 2: Overview of the AR wayfinding guidance system. Images (a-1) through (a-3) show the activation of a landmark touchpoint, bringing up relevant information details. Image (b) shows a user lifting their left hand to activate the interactive 3D map. Images (c-1) and (c-2) show the user activating overt directional guidance. Note: all names and room numbers have been altered and do not correspond to actual locations.

## 4 STUDY 1: USABILITY OF NAVMARKAR

After developing an initial prototype of the NavMarkAR system, we conducted a preliminary study with older adult participants to evaluate the user experience. We were particularly interested in this part of the project to understand how users would interact with the available information and guidance and how positive their reactions would be towards the tool. We encouraged the participants to try out all of the modalities of NavMarkAR, including both visual and auditory cues for landmarks and directional guidance, and the 3D interactive map.

### 4.1 Participants

Six older adult participants were recruited using university e-mail lists. The participants were required to affirm that they had no significant mobility issues, no history of epilepsy or motion sickness, no medical implants, and were unfamiliar with the building in which the study would take place. All of the participants reported that they had little or no prior experience with augmented reality. Three of the participants reported being female, and 3 reported being male.

All participants were between 60–75 years of age (M=68, SD=4.2). The study procedures were approved by the Institutional Review Board (IRB), and all participants provided informed written consent prior to research activities.

### 4.2 Procedure

Our user experience study was conducted in a well-illuminated indoor environment, encompassing one entire floor of a university campus building. We planned six simple routes, each of which was approximately 80 feet in length and required navigating one intersection. The experiment was devised to evaluate the different modalities. The initial task provided only auditory guidance for both directions and landmarks. The second task allowed users to access visual guidance for directions only, and the third was restricted to visual cues for landmarks. During tasks four to six these available modalities were reversed, transitioning from visual cues to only auditory cues; however, in these tasks the users also had access to the 3D interactive map. The study was conducted using a single session for one participant at a time, lasting approximately 1 to 2 hours for each participant. After instructing the participants to put on the HoloLens and familiarize themselves with the device and its interface, we asked them to complete each task in sequence, followed by a post-experiment interview. The interview was semi-structured and each lasted around 30 minutes, covering questions related to the system usability, preference concerning the different guidance modalities, acceptance of the technology, and suggestions for improvements.

### 4.3 Results

*4.3.1 Overall Usability.*

The participants were generally positive about their experience using NavMarkAR. All of the participants were able to successfully complete the six wayfinding tasks using the available task-specific modalities. Participants reported that the navigation guidance was easy to understand, and they appreciated the interactivity of the system. Specific features that participants emphasized for praise during the interviews included the use of virtual touchpoints, the presence of the blue polyhedron to draw attention to landmarks, and the seamless integration of the virtual components with the real environment.

In terms of their guidance preferences, the majority of the participants (83%) expressed a preference for auditory cues in directional guidance, while only one participant preferred the visual cues for this purpose (P3). In contrast, all of the participants preferred the pop-up windows containing text and image for displaying landmark details, rather than using audio recordings to convey this information. One participant emphasized that the visual images they had seen in the landmark descriptions helped them in recalling locations (P2). The pressable virtual touchpoints were a feature that elicited frequent positive participant feedback; one notable comment was that this format allowed users to control what information was displayed and prevented them from being overwhelmed with excessive information (P4). Several participants commented on the clarity and ease-of-use of the map feature. Despite our efforts to reduce over-reliance on the map, one participant noted that they found themselves paying more attention to the map than to the actual surrounding environment (P5).

*4.3.2 Suggested Improvements.*

We asked participants to provide any recommendations that they might have for improving the features of the AR navigation tool. On the basis of this feedback, three main areas were identified for improvement. First, while all participants found the existing directional guidance to be useful, several expressed a desire to receive additional confirmations that they were on a correct route, particularly along lengthy pathways and at intersections where we had not programmed directional cues (P1, P4). They also underscored the importance of having affirmative feedback upon reaching a destinations to reinforce users' confidence (P3).

Second, participants exhibited a strong preference for concise and streamlined guidance during the navigation tasks, emphasizing the value of receiving step-by-step instructions. Some participants felt overwhelmed and distracted by the directional guidance that indicated multiple possible routes (P1). In a related consideration, several participants reported experiencing confusion when they encountered navigational guidance that was intended for a separate, overlapping task-route in our study. In some cases, this resulted in the selection of a touchpoint that sent them in the wrong direction.

Finally, some participants expressed practical concerns about the limited field-of-view of the HoloLens 2 device, which could result in missing some of the available AR landmarks or cues, or losing sight of them if they briefly turned their gaze away. This required the participants to search around visually by turning their heads to try to locate the touchpoints. Additional interface concerns that emerged in the interviews were that some participants perceived the size of the informational text as too small for comfortable reading (P4), and some had trouble managing the display windows, resulting in information being obscured behind other elements or positioned too low in the field of view (P3).

### 4.4 Post-study System Design Improvements

Based on the feedback from Study 1 we made some design changes to improve the NavMarkAR system, adding two new features and several minor enhancements.

*4.4.1 Directional Confirmation Strategies.*

To increase directional confirmation, we implemented on-track and off-track notifications. These touchpoints were strategically placed near intersections that might generate wayfinding confusion, particularly those with multiple pathways, as well as near transition areas leading to different floor levels, and in any passages that extended beyond 50 feet without other guidance. When participants passed by these designated areas, they experienced, as relevant, either a green "on-track" touchpoint affirming the correct trajectory, or a red "off-track" touchpoint signaling an incorrect route (Figure 3). Users then could press these touchpoint buttons and they could receive an auditory cue either reinforcing the current path or prompting them to retrace their steps to the previous landmark or intersection. Moreover, as an additional layer of confirmation, on-screen congratulatory text appeared when users reached the destinations of each wayfinding task. The 3D interactive map feature was not used in subsequent iterations of NavMarkAR. This was done to help to balance the addition of on-track/off-track cues by eliminating an aid that some participants relied on as a substitute for forming their own cognitive maps.

*4.4.2 Simplification of Navigation Guidance.*

We adjusted the system to fix the problem reported in Study 1 where participants encountered guidance intended for other navigation tasks with overlapping routes. In the revised tool, only guidance relevant to the current wayfinding task was visible. In addition, once the user interacted with and closed a particular navigation touchpoint, that touchpoint became invisible so that it would not distract from the upcoming information items. We implemented a command "ResetMap" that allowed users to restore all navigational cues in cases where they felt disoriented or wished to revisit the previous guidance. We also simplified the directional guidance so that it showed only a single route suggestion. Our thinking in making this change, in accordance with the interview data, was that users who activated the directional guidance system were already expressing exhaustion with finding their own way, and would likely be dissatisfied if the directions did not provide a clear, single route suggestion.

*4.4.3 Additional Enhancement.*

We made some additional changes to the NavMarkAR system after Study 1 to enhance the overall user experience. The elevation of interactive elements was adjusted to accommodate users of varied heights, and the size of touchpoints, landmark objects, and textual elements were augmented to improve their visibility. We leveraged the TextMeshPro [75] function in the Unity3D engine to add accessibility adjustments that can be used by individuals with reduced vision. In addition, new notifications were displayed at the forefront of the user's screen, leveraging the eye-tracking capabilities from MRTK plugins, to avoid this information being obscured by other open windows. The RadialView [53] feature of MRTK was incorporated to guarantee that popup windows consistently aligned with the user's field of view. Finally, we minimized the required interaction distance for activating touchpoints. All of these changes were intended to provide a more intuitive and natural feel for diverse users.

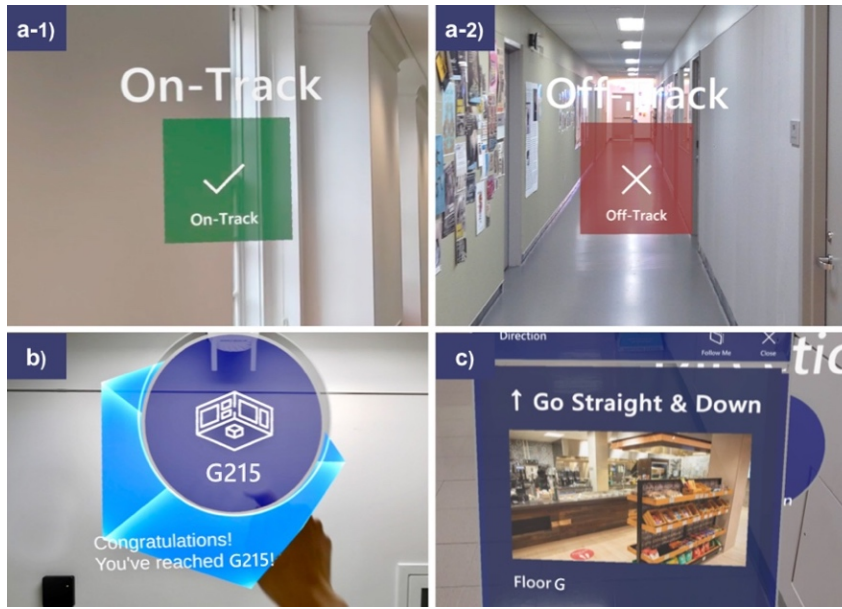

Figure 3: System design improvements after Study 1 included (a-1 and a-2) the Implementation of "on-track" and "off-track" signs, (b) confirmations when users arrive at a destination point, and (c) directional guidance reduced to only one recommended route. Note: all names and room numbers have been altered and do not correspond to actual locations.

## 5 STUDY 2: WAYFINDING PERFORMANCE AND COGNITIVE MAP DEVELOPMENT

After improving the design of NavMarkAR, we conducted a second study to assess the efficacy of the system in enhancing older adults' wayfinding performance and supporting cognitive map development.

### 5.1 Participants

We recruited 32 older adult participants using a targeted convenience sampling method (word-of-mouth and university e-mail lists). Sixteen of these participants were sourced from another ongoing study in our lab. Prior to any research activities, the study procedures received approval from IRB. All participants were informed of the study's goals and requirements and provided written informed consent. The average age of the participants was 70.9 (SD=7.4). The sample skewed toward Female (78%) and White (90%). The participants were well educated, including 11 (34%) with a doctoral degree, 10 (31%) with a professional degree, 7 (22%) with a four-year college degree, and 4 (13%) with some college education (Table 1). We asked the participants to complete the Spatial Anxiety Scale [29] and the Santa Barbara Sense of Direction Scale (SBSOD) [49]. Both of these instruments have been linked with navigational performance and have been found to have good internal consistency. The average score on the Spatial Anxiety Scale was moderate at 49.7 (SD=16.8; possible range 24–120, with higher scores indicating greater anxiety). For the SBSOD, the average score was also moderate at 4.6 (SD=0.9, range 1–7, with higher scores indicating a better sense of direction). These demographic features and scores were not treated as variables in the study, but are presented here to provide an overall snapshot of the participant sample.

Table 1: Participant information

|  | AR | Control | Overall |
|---|---|---|---|
| **Age** | 69.3 (5.7) | 72.6 (8.7) | 70.9 (7.4) |
| **Gender** |  |  |  |
| Women | 11 (68%) | 14 (88%) | 25 (78%) |
| Men | 5 (32%) | 2 (12%) | 7 (22%) |
| **Ethnicity** |  |  |  |
| White | 14 (88%) | 15 (94%) | 29 (90%) |
| Others | 2 (12%) | 1 (6%) | 3 (10%) |
| **Education Level** |  |  |  |
| Doctorate | 5 (31%) | 6 (38%) | 11 (34%) |
| Professional | 5 (31%) | 5 (31%) | 10 (31%) |
| Four-year College | 3 (19%) | 4 (25%) | 7 (22%) |
| Some College | 3 (19%) | 1 (6%) | 4 (13%) |
| **Spatial Anxiety** | 61.1 (9.4) | 38.3 (14.7) | 49.7 (16.8) |
| **SBSOD** | 4.7 (0.9) | 4.6 (1.0) | 4.6 (1.1) |

Notes: Self-reported gender, ethnicity, and education level are shown as n and (%). Other measures are shown as mean and (SD).

## 5.2 Environment and Procedure

The study was conducted on-site in a recently renovated university campus building, which was chosen due to its distinctive and complex architectural design. We created 10 wayfinding tasks, encompassing three floors of the building, using selected interior landmarks as the starting and ending locations (Table 2 and Figure 4). The routes ranged from 120 to 360 feet and incorporated different levels of complexity in regard to the number of intersections, floor changes, and on-route environmental features, which allowed us to evaluate the effectiveness of the navigation system for both easy and difficult wayfinding challenges [20]. The 10 tasks were arranged into two continuous loops (5 tasks per loop), with the next task starting where the previous one ended. Each participant was assigned to a random starting task number for each set; that is, they began at different places in the two loops. We also set up a training area that included a single 60-feet route with one intersection.

The experiment used a factorial design, with a single independent variable of navigation mode (AR-assisted wayfinding vs. non-assisted wayfinding). As the main objective of the study was to examine cognitive map development, we chose the non-AR control condition because it is more effective in spatial learning than a TBT navigation system, focusing primarily on egocentric cues [30, 44]. It also allowed for a more direct assessment of our system, as using other baselines may potentially introduce variables that masked the specific cognitive benefits. The assignment between these two conditions was randomized. All participants were asked to complete the same 10 wayfinding tasks in a randomized order. To assess how participants maintain spatial knowledge after using AR, the experiment group used our AR navigation system for first five trials, but then proceeded without AR for the next five. The control group conducted all

A Landmark-based AR Wayfinding System for Older Adults                                                                                                                  13 ten of the trials without AR guidance. Sessions were conducted for one participant at a time, lasting about 2 to 3 hours in total for each participant.

At the start of the session, participants from the experiment group were first asked to complete a brief training route to become familiar with the HoloLens technology and the navigational aids. This training occurred in a separate location from the wayfinding trials, with a primary focus on familiarizing participants with interacting with the floating buttons and windows as well as ensuring that the sound and brightness settings were comfortable. No performance or user-response data were collected during the training period. Participants from the control group were instead asked to watch a brief informative TED video that was unrelated to the study. Next, the participants began the wayfinding tasks. For each trial, they were instructed to find a particular room in the building as quickly as possible with a typical walking speed. A researcher trailed the participant at a distance of approximately 5 to 6 feet during these wayfinding tasks. If participants were struggling or lost for more than ten minutes, or if they opted to abandon the task, then the researcher would promptly guide them to the end point. Following each of the wayfinding trials, responses to a self-report questionnaire were collected by the researcher, and participants were asked to complete a pointing task and a distance estimation task (these measures are discussed in section 5.3. below).

Once all 10 wayfinding tasks were complete, participants were asked to create a sketch-map, which is a commonly used approach to measure the development of cognitive maps [68]. They were also asked to draw the layout of one entire floor of the building in which the study took place (the same pre-determined floor for all participants), as best as they could from memory. The participants in the AR-aided group were also asked to evaluate the usability and impact of NavMarkAR through the System Usability Scale [35], the MEC Spatial Presence Questionnaire (MEC-SPQ) [58], and the Situational Awareness Rating Technique (SART) [60]. Finally, we conducted a semi-structed short interview to gather insights about participants' experiences using the AR navigation system, suggestions for potential design improvements, and preferences for guidance modalities. The interviews were audio-recorded, transcribed, and analyzed using the flexible coding method [55].

Table 2: The ten wayfinding tasks used in the study

| First Task Loop | | | Second Task Loop | | |
|---|---|---|---|---|---|
| Task Number | Start and End Locations | Changes in Floor Levels | Task Number | Start and End Locations | Changes in Floor Levels |
| 1 | Room G100–Room G200 | Single level | 6 | Room 1300–Room 2300 | Multi-level |
| 2 | Room G200–Room 2000 | Multi-level | 7 | Room 2300–Room 2200 | Single level |
| 3 | Room 2000–Room 2010 | Single level | 8 | Room 2200–Room 2400 | Single level |
| 4 | Room 2010–Room 1100 | Multi-level | 9 | Room 2400–Room G215 | Multi-level |
| 5 | Room 1100–Room G100 | Multi-level | 10 | Room G215–Room 1300 | Multi-level |

Notes: all room numbers have been altered and do not correspond to actual locations.

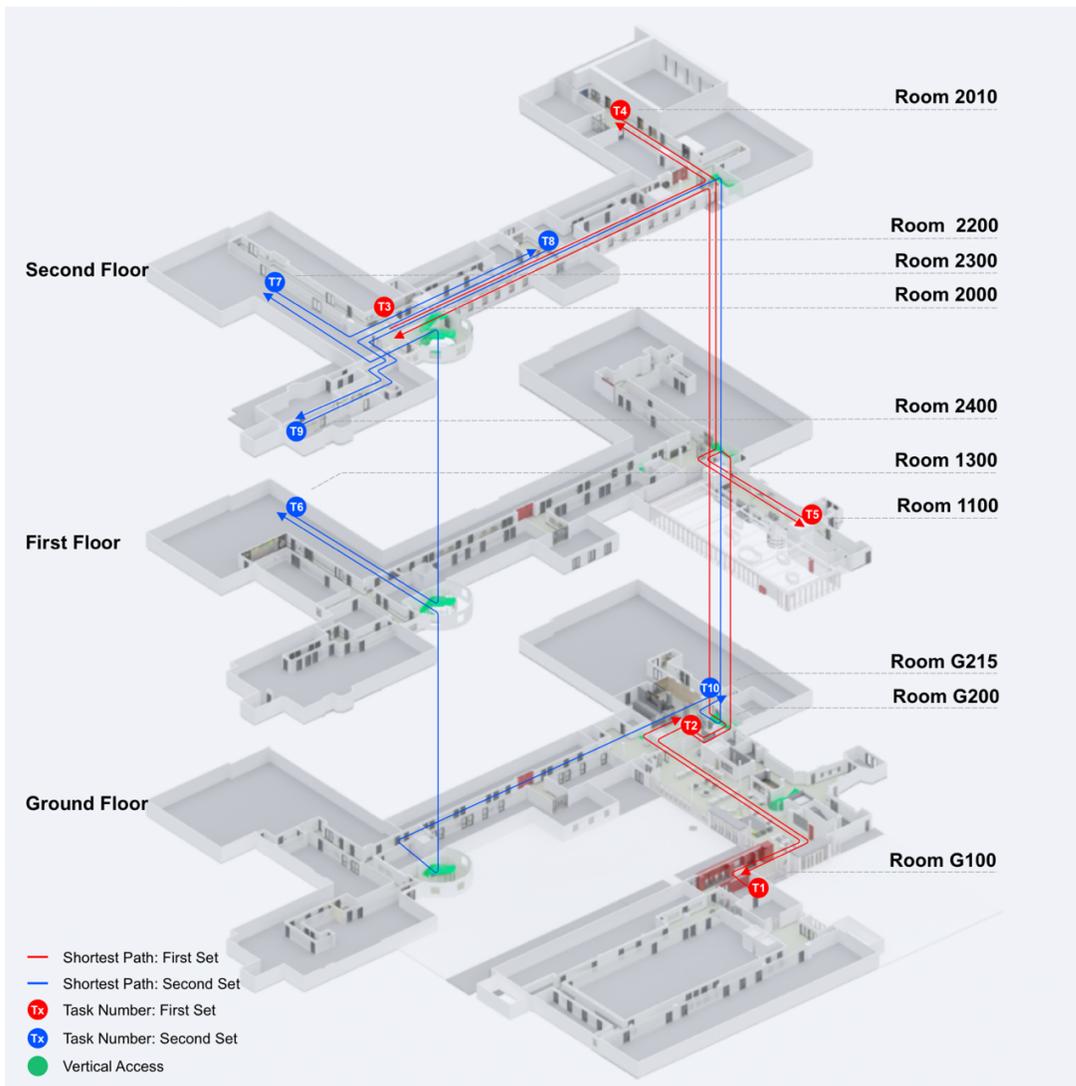

Figure 4: All ten wayfinding tasks as shown on building floorplans.

## 5.3 Measures

We used two standard wayfinding performance metrics: the time required for task completion, and the distance traveled during each task. Participant trajectories on the floorplan were automatically recorded as a set of timestamps in 2-second intervals, using the established localization capacities of our system (Section 3.1), which allowed us to easily collect this data. Additionally, we used stopwatch to manually record the time duration for participants to complete tasks for more accurate measure. For the post-task questionnaires, we used environmental engagement metrics adapted from the Short Form User Engagement Scale [26], which is a 5-point Likert scale that encompasses focused attention, perceived usability, aesthetic appeal, and psychological reward factors associated with one's surroundings. We also assessed participants' cognitive load using the 7-point NASA Task Load Index [64]; and collected their spatial anxiety levels with a format adapted from the 5-point Spielberger State–Trait Anxiety Inventory (STAI) [69]. The adaptions made to these instruments were intended to reduce their complexity since participants were required to complete the questionnaires multiple times during the experiment.

After each trial, participants completed a pointing task, in which they were asked to stand at the route endpoint, facing a designated direction, and then point towards the straight-line direction where they perceived the task starting point to be [77]. The errors were measured by the discrepancy between the actual angles from start to end point and the angles indicated by participants (only horizontal angle was considered in this measure; vertical angle between floors was ignored). For the distance estimation task, participants were asked to estimate the straight-line distance between the start and the end points [57]. We determined distance estimation errors as the discrepancy between the participant estimates and the actual straight-line distance.

The sketch maps were graded for accuracy based on the correct identification of landmarks (3 possible points), route segments (4 possible points), and intersections (4 possible points) [46, 52]. Scores on the SUS instrument, a 10-item, 5-point Likert scale, were normalized to a 0–100 range. The MEC-SPQ scores were averaged across the scale's two dimensions ("self-location" and "possible actions"). The SART scores were summed across the scale's three components ("demand," "supply," and "understanding").

### 5.4 Data Analysis

The R language was used for data analysis. In evaluating the wayfinding performance measures, we noted that some participants either failed or gave up on certain tasks. To address this, we used mixed-effects Cox regression models, with fixed effects of experiment condition for each set of tasks, as well as the random effects of tasks and participants to validate counterbalancing. We then estimated and compared the marginal means of the two experimental conditions. This model allows us to examine how specific factors influence the rate of a particular event happening (e.g., complete a wayfinding task) at a particular point of time, referring to as hazard ratio. The test also reports "survival probability," which is the likelihood that an individual will not experience that event within a period of time, estimating how long they remain in "survival" state before it occurs.

To assess the development of cognitive maps, we fit linear mixed-effects regression models (LMEM). This approach was applied to both pointing task errors and distance estimation errors, incorporating the random effects of task order and participants. For the sketch map scores we used a t-test to evaluate the differences in scores between the AR-aided condition and the non-AR condition.

The NASA TLX and STAI scales were analyzed with LMEM to compare between the experimental and control conditions across all wayfinding tasks. Average scores from MEC-SPQ, SART, USE-SF, and SUS were calculated to provide insights into the overall performance of the AR navigation system.

### 5.5 Results

*5.5.1 Wayfinding Performance.*

For task duration, in the first loop of 5 wayfinding tasks, the AR-assisted participants (M=139.7, SD=51.1) significantly outperformed the control group (M=235.2, SD=174.7), completing the tasks nearly twice as quickly on average ($p < 0.01$). However, the control group showed a markedly lower hazard ratio (HR=0.3), suggesting a 70% decreased likelihood of completing tasks compared to the experiment group (Contrast Estimate (CE)=1.21). For the second loop of 5 wayfinding tasks, there were no significant difference in completion time between the experiment group (M=205.9, SD=157.6) and the control group (M=234.9, SD=173.9, $p$=0.26), and the associated hazard ratio (HR = 0.7) was less distinct than in the first task loop. In these results, it is important to note that two of the non-AR navigators gave up on some of the wayfinding tasks without completing them (a total of 5 tasks between the two participants). The participants indicated that difficulties and frustration in navigating the environments was the main reason for abandoning the tasks.

Regarding distance traveled, in the first loop of wayfinding tasks the AR-assisted participants (M=365.2, SD=127.5) had a significantly shorter travel distance than the control group (M=583.8, SD=481.5) (CE=0.73, $p < 0.01$), further translating into a decreased likelihood to complete tasks (HR=0.48). Findings in the second loop were similar, with a significant difference in the distance traveled between the experiment group (M=524.6, SD=323.3) and control group (M=675.8, SD=452.7) (CE=0.49, HR=0.61, $p < 0.01$). We also found a significant interaction effect between groups and SBSOD scores ($p=0.013$) for distance traveled, indicating that participants who had a good sense of direction benefited more from the AR technology. Example graphs for the task duration and distance traveled in the second task loop are shown in Figure 5.

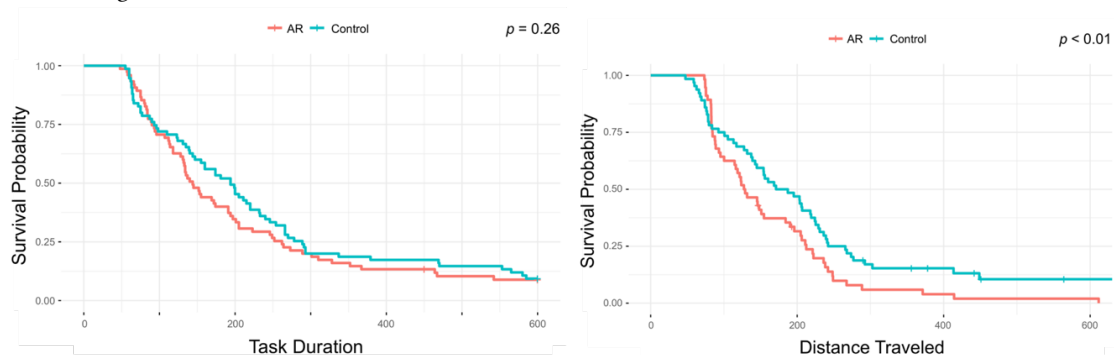

Figure 5: Comparisons of survival probabilities (likelihood to not completing tasks over certain time period) of task duration and distance traversed for the second task loop.

*5.5.2 Cognitive Map Development.*

Results from the linear mixed-effects regression analysis indicated that the AR-aided group exhibited significantly lower directional pointing error compared to the control group, for both the first and second task loops (Table 3 and Figure 6). For the distance estimation task, we capped outliers at the 99th percentile, given that some participant had extremely high estimates. In the first task loop, the discrepancy in distance errors between the AR-aided and control groups was not statistically significant. However, in the second set of tasks, the AR-aided group demonstrated significant lower distance estimation errors compared to the control, indicating that participants were able to maintain cognitive map even after the use of AR navigation system ($p < 0.01$). The t-test analysis for the sketch map scores likewise indicated that participants in the AR-aided group had significantly improved outcomes, with averages of 6.4 vs. 4.4, respectively, out of 11 possible points ($p < 0.01$).

Table 3: Comparisons of performance for cognitive map development

| Task | Task Loop | Experiment Group | Control Group | $p$ |
|---|---|---|---|---|
| Directional Pointing | First Loop | 33.36 (27.85) | 78.6 (55.73) | <0.01** |
|  | Second Loop | 34.61 (31.81) | 67.83 (47.95) | <0.01** |
| Distance Estimation | First Loop | 134.88 (97.68) | 176.02 (133.65) | 0.17 |
|  | Second Loop | 126.03 (100.02) | 184.19 (118.49) | <0.01** |
| Sketch Map |  | 6.41 (2.81) | 4.43 (2.13) | <0.01** |

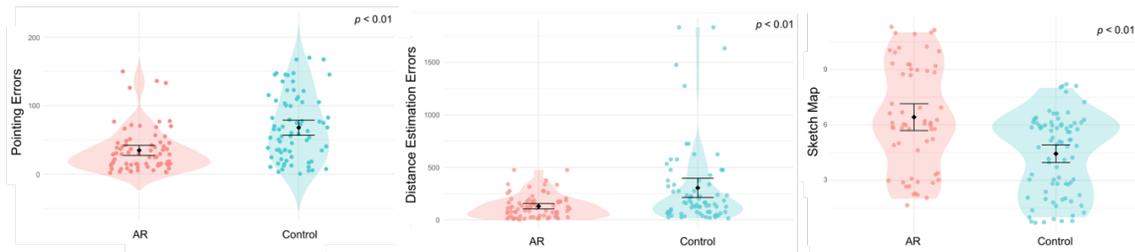

Figure 6: Comparisons of cognitive map development between AR and control group for second task loop; Error bars indicate model-estimated 95% confidence intervals.

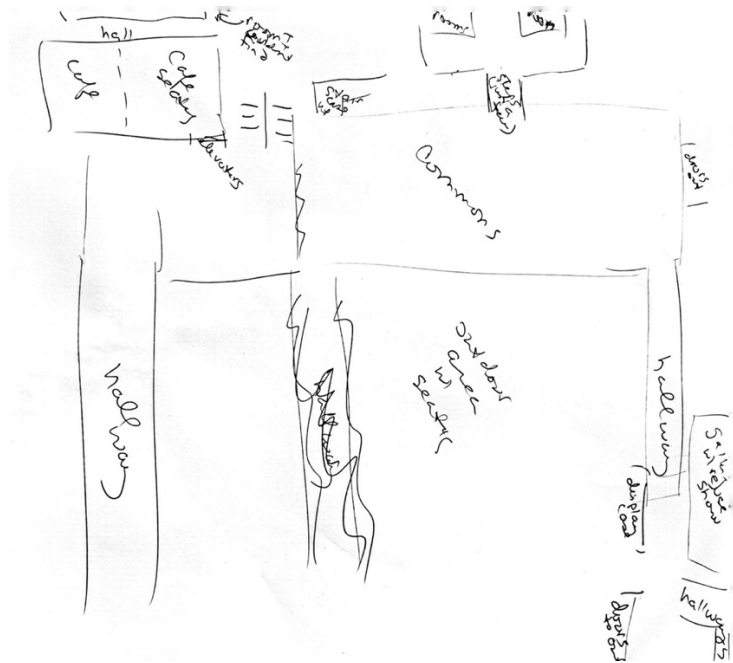

Figure 7: Sketch map from P5: A map with full mark of 11 points that identified landmarks, nodes, and road segments.

### 5.5.3 Cognitive Load and Spatial Anxiety.

We evaluated cognitive load and spatial anxiety associated with the wayfinding system using the NASA TLX and STAI scales, respectively. In the first task loop, the AR-aided group reported significantly lower cognitive loads (M=1.58, SD=0.58) compared to the control group (M=2.16, SD=1.16, p=0.02). However, in the second loop, both groups demonstrated comparable mental workloads ($p$=0.25). Regarding spatial anxiety, there were no significant differences found between the two groups in either task loop (first loop $p$=0.44; second loop $p$=0.82).

*5.5.4 User Feedback.*

Participants who were in the AR-assisted group were asked to provide usability ratings and subjective feedback about the tool. Scores from the SUS instrument (M=78.75, SD=16.4; possible range 0–100) indicated that the system was rated as having excellent usability. A majority of participants stated in the interviews that they found the navigational cues easy to understand, suggesting that NavMarkAR effectively directed them through the unfamiliar indoor environment. One participants remarked: "*It would be especially useful in places I've never been before*" (P17). A notable finding is that the route confirmation via tracking touchpoints that we added after Study 1 was noted favorably by multiple participants; for example, P21 stated: "*The tracking touchpoints reinforced my navigation throughout the building, and it was reassuring.*" The directional cues also received positive feedback for their clarity and straightforwardness. While most participants appreciated the comprehensive integration of navigation guidance, some had specific preferences, focusing on the benefits of either the auditory directions (P8, P10), or the visual route previews (P15, P22). We take this as an indication that the tool was successfully addressing multiple individual preferences. Regarding the landmark information displays, many participants found these cues to be subjectively beneficial in establishing cognitive maps, especially for differentiating between floor levels. Participant P11 commented overtly that: "*The landmark information is intuitive and helped me build mental map of specific landmarks on different floors.*" Some participants expressed a desire for customization abilities in the tool that would allow them to add personal landmarks such as unique furniture or designs that they observed in the environment: "*Being able to set personal landmarks and navigate on my own, even if I make mistakes, would be useful*" (P13).

Scores from the UES-SF showed that the AR-guided participants had high levels of environmental engagement (M=4.06, SD=0.67; possible range 0–5); and scores from the MEC-SPQ (M=7.98, SD=2.33; possible range 1–10) indicated that participants felt highly immersed in the AR technology, with a smooth merger between the virtual and physical realms. The SART scores (M=21.57, SD=4.43; possible range 3–39) similarly indicated a good sense of integration between the AR components and the actual environment, indicating that while participants were engaged with AR, they also remained attentive to external environmental cues such as physical direction signs, objects, and pathways. The interviews reflected the participants' overall positive experience with these technological features—for example, one participant commented that, "*Interacting with the navigation cues was fun, and the AR kept me continuously engaged during the navigation process*" (P19).

## 6 DISCUSSION

This paper reports a two-stage evaluation for NavMarkAR, a novel multimodal AR navigation system targeted toward older adult users. In regard to our Research Question 1, the overall results indicated that the tool was highly successful. Although wayfinding performance was not the primary focus of our study, participants using the AR tool completed navigation tasks faster and traversed less distance compared to the control group. More to the point of our interests, the AR-aided group exhibited significantly higher scores in directional pointing and distance estimation, and were more accurate in drawing sketches of the environment. This indicated that NavMarkAR's features effectively aided in the development of cognitive maps in indoor environments. Our findings contrast with much of the prior literature on TBT navigational aids, as researchers have found that the use of such aids impedes or is irrelevant to cognitive map formation [30, 34, 44]. One possible explanation for this discrepancy is that our AR system maintained a manageable level of uncertainty and encouraged users to more actively engage with the physical environment compared to TBT aids. The use of a landmark-based system and overview map in our navigational aid, rather than simple TBT directions, was intended to promote the acquisition of allocentric knowledge. It is particularly notable that in the second task loop, where our experiment group no longer used AR, the participants nonetheless maintained superior cognitive mapping performance, indicating enduring benefits of AR after immediate use. However, other specific outcome elements, such as task duration and reduced mental workload, did not persist when the experimental group stopped using AR.

In regard to our Research Question 2, participants reported generally positive usability ratings and subjective perceptions of the tool's value. We found that providing too much direct guidance could be a liability, as participants expressed appreciation for the empowering aspect of being able to select when and how much navigational information to receive. We therefore suggest that it is important to provide a good balance between features intended to prompt environmental exploration for cognitive map development vs. features that can provide clear and specific TBT directions when needed. Participants appreciated the availability of different guidance modalities, with different individuals preferring different modalities for various types of content. The main challenges noted were of a technological nature, including participants losing track of the position of AR elements due to the smartglasses' limited field-of-view, errors in receiving guidance intended for other navigation tasks, difficulty in reading some of the informational text due to its small size, and difficulty in adjusting the visible AR windows. Fortunately, most of these challenges can be addressed through improved design, as we attempted to accomplish by improving our prototype between Study 1 and Study 2.

### 6.1 Implications for Multimodal Interactions

Participants found interacting with navigation touchpoints to be intuitive and natural, and reported a high degree of fluid immersion and engagement with AR elements, an experience that has previously been correlated with positive learning outcomes [27]. While there were variabilities in participant preferences for auditory vs. visual feedback, we found that auditory cues were most often preferred for directional and on-track information, and visual cues were preferred for the landmark information. The preference for visual cues could be attributed to the inherent visibility and geometrical saliency of prominent environmental features, and may be linked to the development of allocentric perspectives [6, 36, 50]. In contrast, the preference of auditory cues for immediate directional instruction may align with a desire for simple and non-intrusive assistance with low cognitive load [24, 73]. In future iterations of NavMarkAR, we plan to add more personalization options that will allow participants to adjust the information modality based on their individual preferences (Section 6.3).

### 6.2 Implications for Simplicity and Controllability of Navigational Aids

Participants expressed a strong preference for simplicity in navigational guidance, as well as the flexibility to decide when such guidance would be presented. These preferences are likely related to reducing the cognitive load that wayfinding activities impose upon users [43]. Our results notably indicated that the older adult participants drew more accurate sketch-maps when they received streamlined directional cues, compared to being provided with multiple route options. This contradicts previous studies in which freedom of choice was found to aid spatial learning in younger populations [21, 25]. We also observed that while participants appreciated the potential of the 3D interactive map to enhance environmental awareness, they also characterized it as leading to mental overload. If the fatigue or cognitive-overload response hindered the translation of spatial information into cognitive map formation, or detracted from environmental engagement, that could explain why NavMarkAR's more open-ended features were associated with decreased spatial learning in the older adult population. We suggest that the competing goals of simplicity and engagement need to be carefully balanced, and additional empirical research would be valuable to better understand the exact "sweet spot" for ease-of-use vs. spatial skills development. Features that are not included in our current version of NavMarkAR, such as virtual highlights of existing directional signs in the physical environment, or the addition of a compass indicating North, might help to simultaneously improve both ease of use and engagement with the physical surroundings [32].

### 6.3 Design Recommendations

Based on the findings of our study, we suggest the following three primary guidelines for designing future AR navigation support systems that can enhance cognitive map formation, particularly for older adults and those with MCI. First, the system should seek ways to balance and simultaneously enhance the goals of simplicity vs. environmental engagement. Egocentric-based navigation is often desired by users, and our results indicate that an overemphasis on promoting allocentric spatial learning can cause older adults to feel overwhelmed, with diminishing benefits for cognitive map formation. However, in contrast, when egocentric directions are the sole modality, they may leave little room to foster users' spatial skills [40, 45]. Thus, designers should seek to ensure that engagement opportunities are highlighted in the system, while still making egocentric directions available when needed. Whenever possible, directions should include pointers to actual environmental objects, such as signs, landmarks, and hallways, both in visual feedback and in audio directions. Such approaches will help in creating AR navigation aids that are engaging and useful, both for immediate navigation goals and for the development of spatial understanding, decision-making, and problem-solving abilities.

Second, such systems should emphasize customization to allow users to personalize the navigation guidance to their specific needs. As people have varied preferences for different types of visual and auditory information, and may have various needs during different navigational tasks, the system should offer as much flexibility as possible for the user to choose their preferred modalities and select the level of assistance that they require. For example, sliders could allow users to select the overall amount of guidance that they want to receive, the balance between landmark/mapping information vs. TBT directions, and the balance between audio vs. textual information. Such flexibility will promote acceptance of the technology and empower users to engage with its available skill-building components in times and places that best suit their needs. Additional customization features may be useful, such as the ability for users to define their own landmarks [14] or add personal narratives and notes to a landmark's information display.

Finally, AR navigation systems should consider adaptive interfaces that can change the content display and feedback methods based on user interactions. For example, if users have a prolonged gaze duration or repeated pauses at intersections, then the system can simplify its display or moderate the information density to reduce cognitive load and enhance the effectiveness of the navigational aids. The system should incorporate as much as possible accessibility features to address the diverse minor disabilities common among older adults; this could potentially include enabling voice commands for those who find gestures challenging, offering high-contrast visuals for individuals with vision impairments, and providing tactile feedback for those with hearing difficulties. For any such adaptive technologies, it is very important to provide the user with the ability to deactivate the features if desired, since some users respond negatively to automatic adjustments.

## 7 LIMITATIONS AND FUTURE WORK

While this study provided valuable new knowledge about the cognitive implications of AR navigation tools for older adults, it does have several noteworthy limitations. First, the participant sample skewed strongly toward female, white, and highly educated older adults. Such demographic factors may have an influence on how individuals relate to technology and their navigation needs. Future research should seek to include participants with more diverse demographic attributes to enhance the generalizability of findings. Second, we encountered several technical malfunctions during our research, some of which required restarting the navigation app from the origin point, leading to data loss for participants' trajectories. This required the two participants to re-do some of the navigational tasks, which might have influenced their wayfinding performance and other response measures. In these cases, we discarded the original performance times for the tasks where glitched occurred, and replaced them with the times for the repeated tasks. To prevent hardware malfunctions, we suggest that future studies integrate a computer-vision-based indoor positioning system alongside spatial mapping data to enhance system stability. Finally, two of our participants opted out of completing some of the wayfinding tasks, and two participants opted out of completing the end-of session map-sketch tasks. These exclusions decrease the reliability of the statistical analysis. The primary reason for participant withdraw seemed to be fatigue, which might be alleviated by shortening the duration of the experiment sessions.

## 8 CONCLUSION

In this paper, we presented NavMarkAR, a novel landmark-based AR navigation system designed for older adults, aiming to enhance their wayfinding performance and spatial skills maintenance. Our empirical research on this system's implementation assessed older adult participants' wayfinding performance, cognitive map development, and technology preferences, comparing outcomes for NavMarkAR users against a control group. The study findings made three major contributions to the field. First, our results demonstrated that multimodal AR navigation guidance can be highly effective for helping older adults to reach their destinations. Second, the landmark-based design was effective in assisting the older adult participants to acquire allocentric knowledge in indoor environments. Finally, our surveys and interviews revealed important insights into older adults' outlooks and needs in the context of system usability, leading to a set of specific design recommendations for AR navigation tools. We look forward to seeing how future research will expand upon the current work and continue to advance this engaging and useful technology.

## REFERENCES

[1] Aaron L. Gardony, Tad T. Brunyé, Caroline R. Mahoney, and Holly A. Taylor. 2013. How navigational aids impair spatial memory: evidence for divided attention. *Spatial Cognition & Computation* 13, 4 (2013), 319–350. https://doi.org/10.1080/13875868.2013.792821

[2] Adam W. Lester, Scott D. Moffat, Jan M. Wiener, Carol A. Barnes, and Thomas Wolbers. 2017. The aging navigational system. *Neuron* 95, 5 (2017), 1019–1035. https://doi.org/10.1016/j.neuron.2017.06.037

[3] Alan L. Liu, Harlan Hile, Henry Kautz, Gaetano Borriello, Pat A. Brown, Mark Harniss, and Kurt Johnson. 2006. Indoor wayfinding: Developing a functional interface for individuals with cognitive impairments. In *Proceedings of the 8th International ACM SIGACCESS Conference on Computers and Accessibility*, 95–102. https://doi.org/10.1145/1168987.1169005

[4] Angélique Montuwy, Béatrice Cahour, and Aurélie Dommes. 2019. Using Sensory wearable devices to navigate the city: Effectiveness and user experience in older pedestrians. *Multimodal Technologies and Interaction* 3, 1 (2019), 17. https://doi.org/10.3390/mti3010017

[5] Annina Brügger, Kai-Florian Richter, and Sara Irina Fabrikant. 2019. How does navigation system behavior influence human behavior? *Cognitive Research: Principles and Implications* 4 (2019), 1–22. https://doi.org/10.1186/s41235-019-0156-5

[6] Arianna Rinaldi, Elvira De Leonibus, Alessandra Cifra, Giulia Torromino, Elisa Minicocci, Elisa De Sanctis, Rosa María López-Pedrajas, Alberto Oliverio, and Andrea Mele. 2020. Flexible use of allocentric and egocentric spatial memories activates differential neural networks in mice. *Scientific Reports* 10, 1 (2020), 11338. https://doi.org/10.1038/s41598-020-68025-y

[7] Aryan Firouzian, Yukitoshi Kashimoto, Zeeshan Asghar, Niina Keranen, Goshiro Yamamoto, and Petri Pulli. 2017. Twinkle megane: Near-eye led indicators on glasses for simple and smart navigation in daily life. In *eHealth 360°: Lecture Notes of the Institute for Computer Sciences, Social Informatics and Telecommunications Engineering* 181. Springer International Publishing. https://doi.org/10.1007/978-3-319-49655-9_3

[8] Arzu Çöltekin, Ian Lochhead, Marguerite Madden, Sidonie Christophe, Alexandre Devaux, Christopher Pettit, Oliver Lock et al. 2020. Extended reality in spatial sciences: A review of research challenges and future directions. *ISPRS International Journal of Geo-Information* 9, 7 (2020), 439. https://doi.org/10.3390/ijgi9070439

[9] Ashlynn M. Keller, Holly A. Taylor, and Tad T. Brunyé. 2020. Uncertainty promotes information-seeking actions, but what information? *Cognitive Research: Principles and Implications* 5 (2020), 1–17. https://doi.org/10.1186/s41235-020-00245-2

[10] Bing Liu, Linfang Ding, and Liqiu Meng. 2021. Spatial knowledge acquisition with virtual semantic landmarks in mixed reality-based indoor navigation. *Cartography and Geographic Information Science* 48, 4 (2021), 305-319. https://doi.org/10.1080/15230406.2021.1908171

[11] Bingjie Cheng, Anna Wunderlich, Klaus Gramann, Enru Lin, and Sara I. Fabrikant. 2022. The effect of landmark visualization in mobile maps on brain activity during navigation: A virtual reality study. *Frontiers in Virtual Reality* 3 (2022). https://doi.org/10.3389/frvir.2022.981625

[12] Charles Spence. 2020. Senses of place: Architectural design for the multisensory mind. *Cognitive Research: Principles and Implications* 5, 1 (2020), 46. https://doi.org/10.1186/s41235-020-00243-4

[13] Christina Bauer, Manuel Müller, and Bernd Ludwig. 2016. Indoor pedestrian navigation systems. In *Proceedings of the 15th International Conference on Mobile and Ubiquitous Multimedia (MUM '16)*. https://doi.org/10.1145/3012709.3012728

[14] Clark C. Presson and Daniel R. Montello. 1988. Points of reference in spatial cognition: stalking the elusive landmark. *British Journal of Developmental Psychology* 6, 4 (1988), 378–381. https://doi.org/10.1111/j.2044-835X.1988.tb01113.x

[15] Deetje Iggena, Sein Jeung, Patrizia M. Maier, Christoph J. Ploner, Klaus Gramann, and Carsten Finke. 2023. Multisensory input modulates memory-guided spatial navigation in humans. *Communications Biology* 6, 1 (2023), 1167. https://doi.org/10.1038/s42003-023-05522-6

[16] Demet Yesiltepe, Ruth C. Dalton, and Ayse O. Torun. 2021. Landmarks in wayfinding: A review of the existing literature. *Cognitive Processing* 22 (2021), 369–410. https://doi.org/10.1007/s10339-021-01012-x

[17] Desirée Colombo, Silvia Serino, Cosimo Tuena, Elisa Pedroli, Antonios Dakanalis, Pietro Cipresso, and Giuseppe Riva. 2017. Egocentric and allocentric spatial reference frames in aging: A systematic review. *Neuroscience & Biobehavioral Reviews* 80 (2017), 605–621. https://doi.org/10.1016/j.neubiorev.2017.07.012

[18] Dongkai Fan and Ping Shi. 2010. Improvement of Dijkstra's algorithm and its application in route planning. In *2010 Seventh International Conference on Fuzzy Systems and Knowledge Discovery*, 1901-1904. IEEE. 10.1109/FSKD.2010.5569452

[19] Edward C. Tolman. 1948. Cognitive maps in rats and men. *Psychological Review* 55, 4 (1948), 189–208. https://doi.org/10.1037/h0061626

[20] Edward Slone, Ford Burles, Keith Robinson, Richard M. Levy, and Giuseppe Iaria. 2015. Floor plan connectivity influences wayfinding performance in virtual environments. *Environment and Behavior* 47, 9 (2015), 1024-1053. https://doi.org/10.1177/0013916514533318

[21] Elizabeth R. Chrastil and William H. Warren. 2015. Active and passive spatial learning in human navigation: acquisition of graph knowledge. *Journal of Experimental Psychology: Learning, Memory, and Cognition* 41, 4 (2015), 1162–1178. https://doi.org/10.1037/xlm0000082

[22] Eva Nuhn and Sabine Timpf. 2017. Personal dimensions of landmarks. In *Societal Geo-innovation: Selected papers of the 20th AGILE conference on Geographic Information Science*. Springer International Publishing, 129-143. https://doi.org/10.1007/978-3-319-56759-4_8

[23] František Duchoň, Andrej Babinec, Michal Kajan, Peter Beňo, Martin Florek, Tomáš Fico, and Ladislav Jurišica. 2014. Path planning with modified a star algorithm for a mobile robot. *Procedia Engineering* 96 (2014), 59–69. https://doi.org/10.1016/j.proeng.2014.12.098

[24] Gregory D. Clemenson, Antonella Maselli, Alexander J. Fiannaca, Amos Miller, and Mar Gonzalez-Franco. 2021. Rethinking GPS navigation: Creating cognitive maps through auditory clues. *Scientific Reports* 11, 1 (2021), 7764. https://doi.org/10.1038/s41598-021-87148-4

[25] Haosheng Huang, Thomas Mathis, and Robert Weibel. 2022. Choose your own route–supporting pedestrian navigation without restricting the user to a predefined route. *Cartography and Geographic Information Science* 49, 2 (2022), 95-114. https://doi.org/10.1080/15230406.2021.1983731

[26] Heather L. O'Brien, Paul Cairns, and Mark Hall. 2018. A practical approach to measuring user engagement with the refined user engagement scale (UES) and new UES short form. *International Journal of Human-Computer Studies* 112 (2018), 28-39. https://doi.org/10.1016/j.ijhcs.2018.01.004

[27] Heather L. O'Brien. 2016. Theoretical perspectives on user engagement. In *Why Engagement Matters: Cross-disciplinary Perspectives of User Engagement in Digital Media*. 1-26. https://doi.org/10.1007/978-3-319-27446-1_1

[28] HoloLens 2. 2023. https://www.microsoft.com/en-us/hololens/

[29] Ian M. Lyons, Gerardo Ramirez, Erin A. Maloney, Danielle N. Rendina, Susan C. Levine, and Sian L. Beilock. 2018. Spatial Anxiety: A novel questionnaire with subscales for measuring three aspects of spatial anxiety. *Journal of Numerical Cognition* 4, 3 (2018), 526-553. https://doi.org/10.5964/jnc.v4i3.154

[30] Ian T. Ruginski, Sarah H. Creem-Regehr, Jeanine K. Stefanucci, and Elizabeth Cashdan. 2019. GPS use negatively affects environmental learning through spatial transformation abilities. *Journal of Environmental Psychology* 64 (2019), 12–20. https://doi.org/10.1016/j.jenvp.2019.05.001

[31] Ineke J. M. van der Ham and Michiel H. G. Claessen. 2020. How age relates to spatial navigation performance: Functional and methodological considerations. *Ageing Research Reviews* 58 (2020), 101020. https://doi.org/10.1016/j.arr.2020.101020

[32] Jeanine K. Stefanucci, David Brickler, Hunter C. Finney, Emi Wilson, Trafton Drew, and Sarah H. Creem-Regehr. 2022. Effects of simulated augmented reality cueing in a virtual navigation task. *Frontiers in Virtual Reality* 3 (2022). https://doi.org/10.3389/frvir.2022.971310

[33] Jie Lu, Yu Han, Yunfei Xin, Kang Yue, and Yue Liu. 2021. Possibilities for designing enhancing spatial knowledge acquirements navigator: A user study on the role of different contributors in impairing human spatial memory during navigation. In *Extended Abstracts of the 2021 CHI Conference on Human Factors in Computing Systems (CHI EA '21)*. Association for Computing Machinery, New York, NY, USA. https://doi.org/10.1145/3411763.3451641

[34] Jinyue Zhang, Xiaolu Xia, Ruiqi Liu, and Nan Li. 2021. Enhancing human indoor cognitive map development and wayfinding performance with immersive augmented reality-based navigation systems. *Advanced Engineering Informatics* 50 (2021), 101432. https://doi.org/10.1016/j.aei.2021.101432

[35] John Brooke. 1996. SUS: A "quick and dirty" usability. *Usability Evaluation in Industry* 189, 3 (1996), 189-194. Taylor & Francis. ISBN: 0748404600.

[36] Joost Wegman, Anna Tyborowska, and Gabriele Janzen. 2014. Encoding and retrieval of landmark-related spatial cues during navigation: An fMRI study. *Hippocampus* 24, 7 (2014), 853–868. https://doi.org/10.1002/hipo.22275

[37] Juan M. D. Delgado, Lukumon Oyedele, Peter Demian, and Thomas Beach. 2020. A research agenda for augmented and virtual reality in architecture, engineering and construction. *Advanced Engineering Informatics* 45 (2020), 101122. https://doi.org/10.1016/j.aei.2020.101122

[38] Julian Keil, Dennis Edler, and Frank Dickmann. 2019. Preparing the HoloLens for user studies: an augmented reality interface for the spatial adjustment of holographic objects in 3D indoor environments. *KN-Journal of Cartography and Geographic Information* 69, 3 (2019), 205-215. https://doi.org/10.1007/s42489-019-00025-z

[39] Julie A. Jacko. 2009. Human-Computer Interaction. Ambient, Ubiquitous and Intelligent Interaction. In *Proceedings of the 13th International*

A Landmark-based AR Wayfinding System for Older Adults        23

*Conference on Human-Computer Interaction*. Springer International Publishing. https://doi.org/10.1007/978-3-642-02580-8

[40] Katharine S. Willis, Christoph Hölscher, Gregor Wilbertz, and Chao Li. 2009. A comparison of spatial knowledge acquisition with maps and mobile maps. *Computers, Environment and Urban Systems* 33, 2 (2009), 100–110. https://doi.org/10.1016/j.compenvurbsys.2009.01.004

[41] Klara Rinne, Daniel Memmert, and Otmar Bock. 2022. Proficiency of allocentric and egocentric wayfinding: A comparison of schoolchildren with young adults and older adults. *Journal of Navigation* 75, 3 (2022), 528–539. https://doi.org/10.1017/S0373463321000965

[42] Klaus Gramann, Julie Onton, Davide Riccobon, Hermann J. Mueller, Stanislav Bardins, and Scott Makeig. 2010. Human brain dynamics accompanying use of egocentric and allocentric reference frames during navigation. *Journal of Cognitive Neuroscience* 22, 12 (2010), 2836–2849. https://doi.org/10.1162/jocn.2009.21369

[43] Liu Tang and Jia Zhou. 2020. Usability assessment of augmented reality-based pedestrian navigation aid. In *V. Duffy (Ed.), Digital Human Modeling and Applications in Health, Safety, Ergonomics and Risk Management. Posture, Motion and Health. HCII 2020. Lecture Notes in Computer Science*, 12198. Springer International Publishing. https://doi.org/10.1007/978-3-030-49904-4_43

[44] Louisa Dahmani and Véronique D. Bohbot. 2020. Habitual use of GPS negatively impacts spatial memory during self-guided navigation. *Scientific Reports* 10, 1 (2020), 6310. https://doi.org/10.1038/s41598-020-62877-0

[45] Lukáš Hejtmánek, Ivana Oravcová, Jiří Motýl, Jiří Horáček, and Iveta Fajnerová. 2018. Spatial knowledge impairment after GPS guided navigation: Eye-tracking study in a virtual town. *International Journal of Human-Computer Studies* 116 (2018), 15–24. https://doi.org/10.1016/j.ijhcs.2018.04.006

[46] Mark Blades. 1990. The reliability of data collected from sketch maps. *Journal of Environmental Psychology* 10, 4 (1990), 327-339. https://doi.org/10.1016/S0272-4944(05)80032-5

[47] Mark C. Schall Jr., Michelle L. Rusch, John D. Lee, Jeffrey D. Dawson, Geb Thomas, Nazan Aksan, and Matthew Rizzo. 2013. Augmented reality cues and elderly driver hazard perception. *Human Factors* 55, 3 (2013), 643–658. https://doi.org/10.1177/0018720812462029

[48] Marko Radanovic, Kourosh Khoshelham, and Clive Fraser. 2023. Aligning the real and the virtual world: Mixed reality localisation using learning-based 3D–3D model registration. *Advanced Engineering Informatics* 56 (2023), 101960. https://doi.org/10.1016/j.aei.2023.101960

[49] Mary Hegarty, Anthony E. Richardson, Daniel R. Montello, Kristin Lovelace, and Ilavanil Subbiah. 2002. Development of a self-report measure of environmental spatial ability. *Intelligence*. 30, 5 (2002), 425-447. https://doi.org/10.1016/S0160-2896(02)00116-2

[50] Mélanie Cogné, Sophie Auriacombe, Louise Vasa, François Tison, Evelyne Klinger, Hélène Sauzéon, and Pierre-Alain Joseph. Are visual cues helpful for virtual spatial navigation and spatial memory in patients with mild cognitive impairment or Alzheimer's disease? *Neuropsychology* 32, 4 (2018), 385. https://doi.org/10.1037/neu0000435

[51] Melissa E. Meade, John G. Meade, Hélène Sauzeon, and Myra A. Fernandes. 2019. Active navigation in virtual environments benefits spatial memory in older adults. *Brain Sciences* 9, 3 (2019), 47. https://doi.org/10.3390/brainsci9030047

[52] Michael J. O'Neill. 1992. Effects of familiarity and plan complexity on wayfinding in simulated buildings. *Journal of Environmental Psychology* 12, 4 (1992), 319-327. https://doi.org/10.1016/S0272-4944(05)80080-5

[53] Mixed Reality Toolkit 2. 2023. https://learn.microsoft.com/en-us/windows/mixed-reality/mrtk-unity/mrtk2/?view=mrtkunity-2022-05

[54] Neil Burgess. 2006. Spatial memory: How egocentric and allocentric combine. *Trends in Cognitive Sciences* 10, 12 (2006), 551–557. https://doi.org/10.1016/j.tics.2006.10.005

[55] Nicole M. Deterding and Mary C. Waters. 2021. Flexible coding of in-depth interviews: A twenty-first-century approach. *Sociological Methods & Research* 50, 2 (2021), 708-739. https://doi.org/10.1177/0049124118799377

[56] Otmar Bock, Ju-Yi Huang, Oezguer A. Onur, and Daniel Memmert. 2023. The structure of cognitive strategies for wayfinding decisions. *Psychological Research* (2023), 1-11. https://doi.org/10.1007/s00426-023-01863-3

[57] Perry W. Thorndyke. 1981. Distance estimation from cognitive maps. *Cognitive Psychology* 13, 4 (1981), 526-550. https://doi.org/10.1016/0010-0285(81)90019-0

[58] Peter Vorderer, Werner Wirth, Feliz Ribeiro Gouveia, Frank Biocca, Timo Saari, Lutz Jäncke, Saskia Böcking et al. 2004. MEC spatial presence questionnaire. *Retrieved Sept* 18 (2015), 6.

[59] Ramón Hervás, José Bravo, and Jesús Fontecha. 2013. An assistive navigation system based on augmented reality and context awareness for people with mild cognitive impairments. *IEEE Journal of Biomedical and Health Informatics* 18, 1 (2013), 368–374.

[60] Richard M. Taylor. 2017. Situational awareness rating technique (SART): The development of a tool for aircrew systems design. *Situational Awareness*, 111-128. Routledge. ISBN: 9781315087924.

[61] Roberta L. Klatzky. 1998. Allocentric and egocentric spatial representations: Definitions, distinctions, and interconnections. *Spatial cognition: An interdisciplinary approach to representing and processing spatial knowledge*, 1–17. Springer International Publishing. https://doi.org/10.1007/3-540-69342-4_1

[62] Ryan McKendrick, Raja Parasuraman, Rabia Murtza, Alice Formwalt, Wendy Baccus, Martin Paczynski, and Hasan Ayaz. 2016. Into the wild: Neuroergonomic differentiation of hand-held and augmented reality wearable displays during outdoor navigation with functional near infrared spectroscopy. *Frontiers in Human Neuroscience* 10 (2016). https://doi.org/10.3389/fnhum.2016.00216

[63] Saman Jamshidi, Mahnaz Ensafi, and Debajyoti Pat. 2020. Wayfinding in interior environments: An integrative review. *Frontiers in Psychology* 11 (2020). https://doi.org/10.3389/fpsyg.2020.549628

[64] Sandra G. Hart and Lowell E. Staveland. 1988. Development of NASA-TLX (Task Load Index): Results of empirical and theoretical research. *Advances in Psychology* 52 (1988), 139-183. North-Holland. https://doi.org/10.1016/S0166-4115(08)62386-9

[65] Sarah L. Bates and Thomas Wolbers. 2014. How cognitive aging affects multisensory integration of navigational cues. *Neurobiology of Aging* 35, 12 (2014), 2761–2769. https://doi.org/10.1016/j.neurobiolaging.2014.04.003

[66]  Scene Understanding SDK. 2023. https://learn.microsoft.com/en-us/windows/mixed-reality/develop/unity/scene-understanding-sdk
[67]  Seungjun Kim and Anind K. Dey. 2009. Simulated augmented reality windshield display as a cognitive mapping aid for elder driver navigation. In *Proceedings of the SIGCHI Conference on Human Factors in Computing Systems*, 133–142. https://doi.org/10.1145/1518701.1518724
[68]  Shannon D. Moeser. 1988. Cognitive mapping in a complex building. *Environment and Behavior* 20, 1 (1988), 21-49. https://doi.org/10.1177/0013916588201002
[69]  Theresa M. Marteau and Hilary Bekker. 1992. The development of a six-item short-form of the state scale of the Spielberger State—Trait Anxiety Inventory (STAI). *British Journal of Clinical Psychology* 31, 3 (1992), 301-306. https://doi.org/10.1111/j.2044-8260.1992.tb00997.x
[70]  Thiago D. Oliveira de Araujo, Claudio G. Resque dos Santos, Romario S. do Amor Divino Lima, and Bianchi S. Meiguins. 2019. A model to support fluid transitions between environments for mobile augmented reality applications. *Sensors* 19 (2019), 4254. https://doi.org/10.3390/s19194254
[71]  Toru Ishikawa. 2019. Satellite navigation and geospatial awareness: Long-term effects of using navigation tools on wayfinding and spatial orientation. *The Professional Geographer*, 71, 2 (2019), 197–209. https://doi.org/10.1080/00330124.2018.1479970
[72]  Toru Ishikawa. 2020. Human Spatial Cognition and Experience: Mind in the World, World in the Mind. Routledge. https://doi.org/10.4324/9781351251297
[73]  Tracey Wallace and John T. Morris. 2018. Development and testing of EyeRemember: A memory aid app for wearables for people with brain injury. In *Computers Helping People with Special Needs: 16th International Conference*, 493–500. Springer International Publishing. https://doi.org/10.1007/978-3-319-94274-2_73
[74]  Umair Rehman and Shi Cao. 2016. Augmented-reality-based indoor navigation: A comparative analysis of handheld devices versus google glass. *IEEE Transactions on Human-Machine Systems* 47, 1 (2016), 140-151. https://doi.org/10.1109/THMS.2016.2620106
[75]  Unity Real-Time Development Platform. 2023. https://unity.com/
[76]  Vanessa Huston and Kai Hamburger. 2023. Navigation aid use and human wayfinding: How to engage people in active spatial learning. *KI - Künstliche Intelligenz* (2023). https://doi.org/10.1007/s13218-023-00799-5
[77]  Veronica Muffato, Chiara Meneghetti, and Rossana De Beni. 2020. The role of visuo-spatial abilities in environment learning from maps and navigation over the adult lifespan. *British Journal of Psychology* 111, 1 (2020), 70–91. https://doi.org/10.1111/bjop.12384
[78]  Weihua Dong, Yulin Wu, Tong Qin, Xinran Bian, Yan Zhao, Yanrou He, Yawei Xu, and Cheng Yu. 2021. What is the difference between augmented reality and 2D navigation electronic maps in pedestrian wayfinding? *Cartography and Geographic Information Science* 48, 3 (2021), 225-240. https://doi.org/10.1080/15230406.2021.1871646
[79]  Yuhang Zhao, Elizabeth Kupferstein, Hathaitorn Rojnirun, Leah Findlater, and Shiri Azenkot. 2020. The effectiveness of visual and audio wayfinding guidance on smartglasses for people with low vision. In *Proceedings of the 2020 CHI Conference on Human Factors in Computing Systems (CHI '20)*. https://doi.org/10.1145/3313831.3376516
[80]  Zhanat Makhataeva, Tolegen Akhmetov, and Huseyin Atakan Varol. 2023. Augmented reality for cognitive impairments. *Springer Handbook of Augmented Reality*, 765–793. Springer International Publishing. https://doi.org/10.1007/978-3-030-67822-7_31